# Modelling of Human Glottis in VLSI for Low Power Architectures

**Nikhil Raj[1] and R.K.Sharma[2]**

**[1,2]Electronics and Communication Engineering Department, National Institute of Technology Kurukshetra, Haryana, 136119, India**

## Abstract

The Glottal Source is an important component of voice as it can be considered as the excitation signal to the voice apparatus. Nowadays, new techniques of speech processing such as speech recognition and speech synthesis use the glottal closure and opening instants. Current models of the glottal waves derive their shape from approximate information rather than from exactly measured data. General method concentrate on assessment of the glottis opening using optical, acoustical methods, or on visualization of the larynx position using ultrasound, computer tomography or magnetic resonance imaging techniques. In this work, circuit model of Human Glottis using MOS is designed by exploiting fluid volume velocity to current, fluid pressure to voltage, and linear and nonlinear mechanical impedances to linear and nonlinear electrical impedances. The glottis modeled as current source includes linear, non-linear impedances to represent laminar and turbulent flow respectively, in vocal tract. The MOS modelling and simulation results of glottal circuit has been carried out on BSIM 3v3 model in TSMC 0.18 micrometer technology using ELDO simulator.

## 1. Introduction

Speech synthesis has been a topic of special interest among the researchers. In speech codec and synthetic speech systems, an efficient representation of speech and naturalness of generated speech are important requirements. An emerging approach to improve the naturalness of synthetic speech is based on exploitation of bio-inspired models of speech production. In real human voice production mechanism, the excitation of voiced speech is represented by glottal volume velocity waveform which is generated by the vibrating vocal folds. This excitation signal, referred as glottal wave has naturally attracted interest of researchers in the area of speech synthesis. Glottal wave estimates are needed during vocal fold function analysis, speaker identification, and natural sounding speech synthesis. Artificial models for the glottal source have been used in order to improve the quality of the synthesis. This paper is focused on the idea of utilizing glottal flow pulses extracted directly from voicing source has been proposed. The proposed concept is based on analog (electrical) model of glottis.

Glottis is the space (opening) between the vocal folds and measurement of glottal flow provides benchmark data for voice source models. The lungs and respiratory muscles act as vocal power supply. Voiced speech is produced when air is expelled from the lungs causing the vocal folds to vibrate in non-linear periodic fashion. Such oscillation is approximated by relaxation oscillator. The ejected air stream flows in form of pulses which further get modulated by the vocal tract. In unvoiced speech, sounds are produced by passing the stream of air through a narrow constriction in the vocal tract. The pulses can also arise by making a complete closure, building up pressure behind it, and then followed by an abrupt release. In the first case, a turbulent flow is produced while in the second case, a brief transient excitation occurs. The puffs of air are shaped into sound waves of speech and eventually, radiated from the lips or nose. The periodic signal generated by the vibrating motion of the vocal folds is termed as the glottal flow, glottal volume velocity waveform, or simply the voice source. The rate at which the vocal folds vibrate defines the fundamental frequency of the speech. In normal speech, fundamental frequency changes constantly, providing linguistic information about emotional content, such as differences in speaker mood. In addition, the fundamental frequency pattern determines naturalness of utterance production.

The organization of this paper is as follows. In Section 2, electrical equivalent model of glottis, implemented using MOS transistor, is presented. Section 3 comprises of simulation results and discussion followed by conclusion in section 4.

## 2. Electrical model of Glottis

The vocal tract can be assumed as a non-uniform acoustic tube, with time-varying cross-sectional areas; terminated by the vocal folds at one end, while the other end by lips and nose. Vocal fold vibration produces a periodic interruption of the air flow from the lungs to supraglottal vocal tract based on the principle of Bernoulli Effect. It has been found that at most frequencies of interest, the glottal source has high acoustic impedance compared to the driving point impedance of the vocal tract. Consequently, a current source is used as the electrical analog model that approximates the volume velocity source, i.e. $U_{gl}$ at the glottis. Alternatively, the constriction at the glottis is represented by variable impedance $Z_{gc}(t)$ that serves to model the constrictions created by the opening and closing of the vocal folds in the glottis and thus model turbulent and laminar flow in the vocal tract. The glottal impedance is modulated by a glottal oscillator to model the opening and closing of the vocal folds. The proposed concept is detailed in [1]. Based on above approximations, the electrical equivalent model of glottis is shown in Fig. 1.

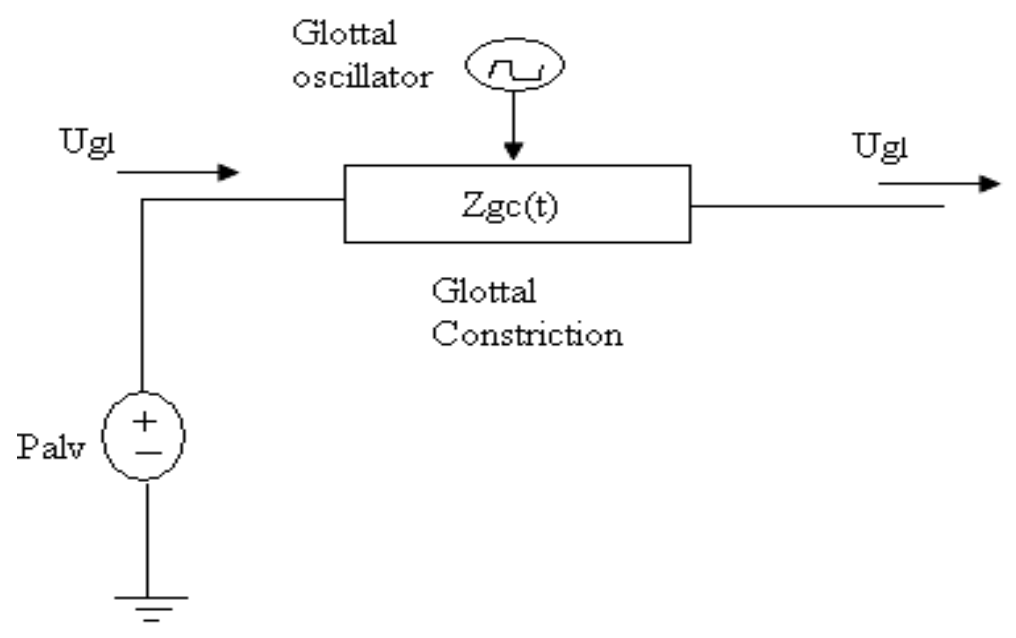

Fig.1 Electrical model of Glottis

### 2.1 Subglottal pressure measurement

Commonly used unit for pressure measurement in speech is $cmH_2O$. The subglottal pressure (lung pressure) is measured in terms of $cmH_2O$, quite similar to that of displacement occurring in U-tube manometer. Yet another device, specialized pressure transducer is also used to measure speech related pressure. Subtenly, Worth and Sakuda used a semiconductor strain gauge device of about 10mm in diameter and 3mm in thickness for lung pressure measurement [2]. This was pasted to palate to sense intraoral pressure during speech. With assumption of system to be linear, using linear regression analysis to translate amplifier output voltage within range to equivalent air pressure; they derived an expression to convert cm $H_2O$ pressure in dc voltage equivalent. The expression takes the form

$$1cmH_2O = 1.27 \times P_{alv} + 5.94 \quad (1)$$

where $P_{alv}$ is the alveolar pressure approximated as dc voltage equivalent. For normal voice, lung pressure ranges from 7-10 cm $H_2O$.

### 2.2 Vocal fold approximation

Human phonation is produced when expiratory air flows through the vocal tract and causes the vocal folds to undergo self-sustained vibration as they exhibit elastic and viscous behavior. The fundamental frequency of human phonation is the fundamental frequency of vocal fold vibration and the intensity of voice is closely related to the amplitude of vocal fold vibration [3]. Different models of vocal fold vibration have been described till now out of which focus is done on multimass model of vocal folds to describe actual behavior of human voicing but their principle of functioning, based on harmonic oscillations, may appear complex. A nonlinear oscillator exhibits a rhythmic burst when excited by an appropriate input stimulus, i.e. lung pressure. Such oscillations can be obtained by using relaxation oscillator [4]. A relaxation oscillator is an oscillator in which a capacitor is charged gradually and then discharged rapidly. The electrical output of a relaxation oscillator is a sawtooth wave in form of current pulses. It modulates the value of $Z_{gc}(t)$ in a periodic fashion to produce a volume velocity waveform $U_{gl}$.

### 2.3 Glottal constriction

Constriction at the glottis can be approximated as a narrow cylindrical duct. Based on posiseuille's law, glottal constriction resistance is modeled as combination of linear and nonlinear resistances to represent losses occurring at the glottis due to laminar and turbulent flow, respectively. A combination of linear and nonlinear resistances is often useful in creating building blocks in electrical models of physical systems. The combination of resistances is in series fashion as shown in Fig. 2. The reason behind is that the upper and lower folds abduct and adduct with a time lag between them [5].

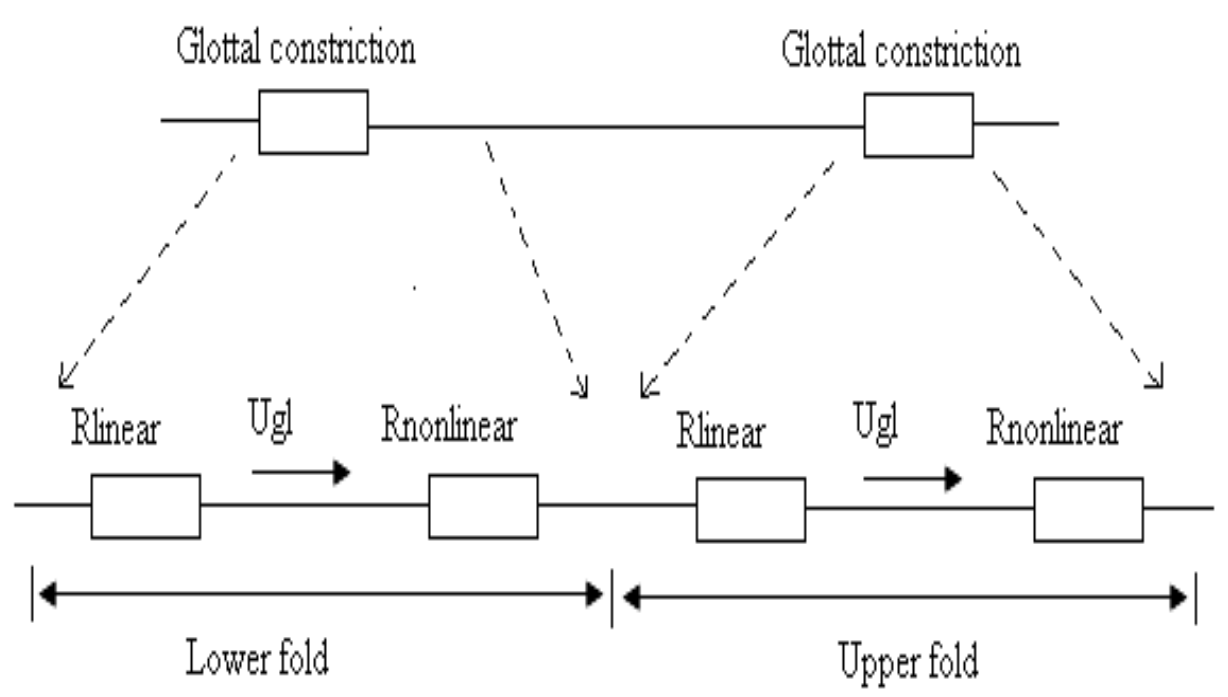

Fig.2 Glottal constrictions to model lower and upper fold

The impedance of each glottal constriction is controlled by a glottal oscillator in a corresponding manner to represent vibration of upper and lower folds.

Electronically tunable resistors are highly versatile circuit elements. They find various applications like in variable gain amplifiers, oscillators, balanced resistive bridges and analog filters. To design such electronically tunable resistors, it is preferably implemented using MOS. As MOS being a four terminal device offers two control parameters that is gate and bulk terminal to control the resistor value. Through electronic tuning of gate terminal voltage of MOS transistor correspondingly electronic control on resistance can be achieved. MOS transistors are generally used for resistor modelling as when it is operated under triode mode behaves as a resistor controlled by its gate terminal voltage. In the past, MOS resistors with approximately linear I-V characteristics were obtained by operating the transistor in the ohmic (triode) region of strong inversion to exploit the resistive nature of the channel. Generally, these approaches were limited by the small ohmic region and its intrinsic non-linearities. Various techniques have been proposed to minimize the nonlinear effects associated with MOS in ohmic strong inversion regime with good results [6]-[8].

For proposed glottis architecture, a MOS resistor is used that does not require triode operation [9]. In addition, this can be applied to produce linear as well as nonlinear resistances. For non-linear I-V characteristics, it uses translinear circuits which incorporate functions like square-root and square. In Fig. 3, the MOS transistor $M_R$ act as a tunable resistor and for tuning its resistive value, a capacitor is connected at its gate terminal. To maintain resistive nature of $M_R$, a feedback network is configured at its gate terminal. As the two OTAs, OTA1 and OTA2 have same inputs, $V_X$ and $V_Y$ connected to their input terminals in alternative fashion, they are biased by the same current source $I_{gm}$ . The potential difference $(V_X - V_Y)$ across the MOS device $M_R$ is sensed and converted into a current $I_{out,gm}$ using a wide linear range OTA [10] for which the output current equation is of the form

$$I_{out,gm} = G_M V_{XY} = G_M \left(V_X - V_Y\right) \qquad (2)$$

where $G_M$ $\left(G_M = I_{gm}/V_L\right)$ is the transconductance of OTA while $I_{gm}$ and $V_L$ are the biasing current and linear range of the OTA, respectively. These two OTAs are configured in conjunction with diode connected transistors M1 and M3 to produce two half-wave rectified currents that are proportional to $|V_{XY}|$, that is, voltage across the source-drain terminals of $M_R$. The rectified output currents get mirrored via M2 and M4 to create a full wave rectified current $I_{in}$. This $I_{in}$ further serves as input current to translinear block and correspondingly output current $I_{out}$ is generated at output end of translinear block as a function of $I_{in}$ i.e. $I_{out} = f(I_{in})$. The translinear block consists of current domain circuits which implement functions like linear, square-root and square. The saturation currents $I_{Xsat}$ and $I_{Ysat}$ of $M_R$ is proportionally replicated by sensing $V_g$, $V_W$, $V_X$ and $V_Y$ on the gate, well, source and drain terminals of $M_R$, buffered via source followers and applying potentials $V_{gX}$ and $V_{gY}$ across the gate-source terminals of transistors $M_X$ and $M_Y$. Transistors M7-M13 serves to compute $I_{Xsat} - I_{Ysat}$ or $I_{Ysat} - I_{Xsat}$ and transistors M14-M17 compare $|I_{Xsat} - I_{Ysat}|$ with a mirrored version of the translinear output current using transistor M6. Any difference between these two currents causes the capacitor $C$ to charge or discharge tuning the gate bias voltage $V_g$ which equilibrates at a point where the two are nearly equal via negative feedback action.

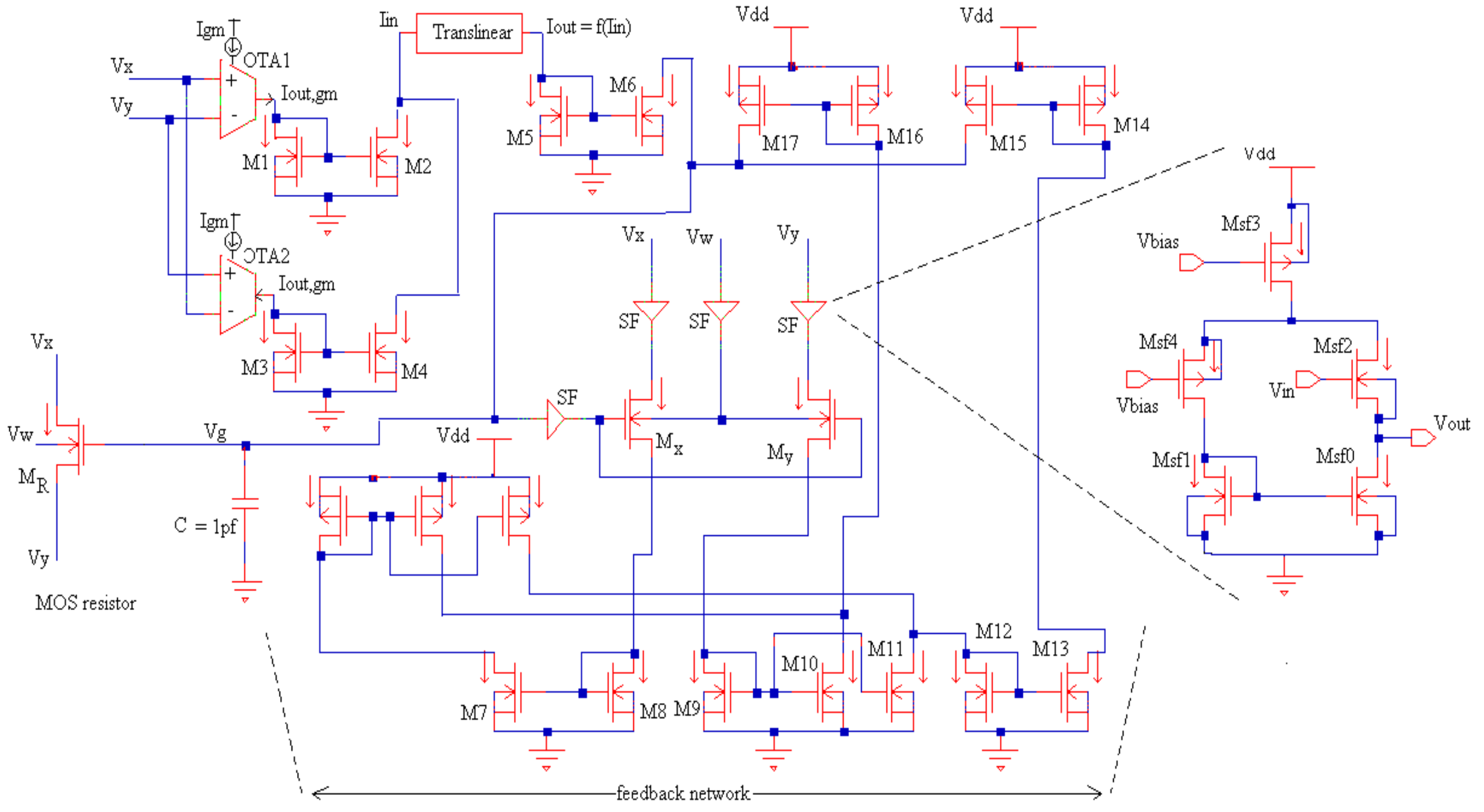

Fig.3 Electronically tunable MOS resistor with feedback circuit

The MOS can resistor can be easily extended to implement a nonlinear resistor which shows behavior of form compressive $\left(I \propto \sqrt{V}\right)$ or expansive $\left(I \propto V^2\right)$ depending on appropriate choice of translinear circuit. For a linear MOS resistor, a translinear circuit with the following input-output relationship $I_{out} = I_{in}$ is used. Likewise, for compressive resistor and for expansive resistor a translinear circuit with relation $I_{out} = \sqrt{I_{in} I_{ref}}$ and $I_{out} = I_{in}^2 / I_{ref}$ is used respectively, where $I_{ref}$ is a reference current.

To overcome loading effect on terminals of MOS transistor $M_R$, source follower is employed shown within dotted lines marked as SF. The source follower has the capability to source and sink large output currents. Its primary use is to buffer signals and provide low output impedance to drive resistive loads while, at the same time handle large output voltage swing and obtain low harmonic distortions. Traditional source have load drive capability limited to the quiescent current in the buffer. In addition traditional source followers require too much power for many applications. To reduce power dissipation (and area), composite source follower is used. The composite source follower comprises a current source, Msf3 configured to provide a (relatively) constant current to the rest of the circuit, a source follower NMOS (Msf0, Msf2) configured to receive an input signal, a folded cascade device Msf4 connected to sense the drain current of the source follower, and a current mirror device Msf1 connected to multiply the sensed drain current for application to an output load connected at the source follower output. By adjusting $W/L$ ratio of transistor Msf0 with respect to Msf1, a four-fold increase in transconductance is obtained which enables perfect tracking of input by output having no level shift problem generally found in common voltage buffers.

## 3. Results and Discussion

As discussed earlier, glottal constriction is a combination of two series connected resistances to model lower and upper vocal folds under laminar and turbulent airflow. With such approximation, the lower fold is modeled as combination of linear and square-root circuit while the upper fold is modeled as combination of linear and square circuit, shown in Fig. 4. The oscillator is provided as an input current pulse which also controls bias current of OTAs used in tunable MOS resistor architecture. As oscillation is approximated by relaxation oscillator, the output is sawtooth wave as current pulse where its rise is kept much higher than that of fall time. For simulation purpose, the pulse duration is kept about 8ms to model vibration of fold in adult man. The time lag of 1ms is kept between successive oscillator input to model lower and upper vocal fold oscillations.

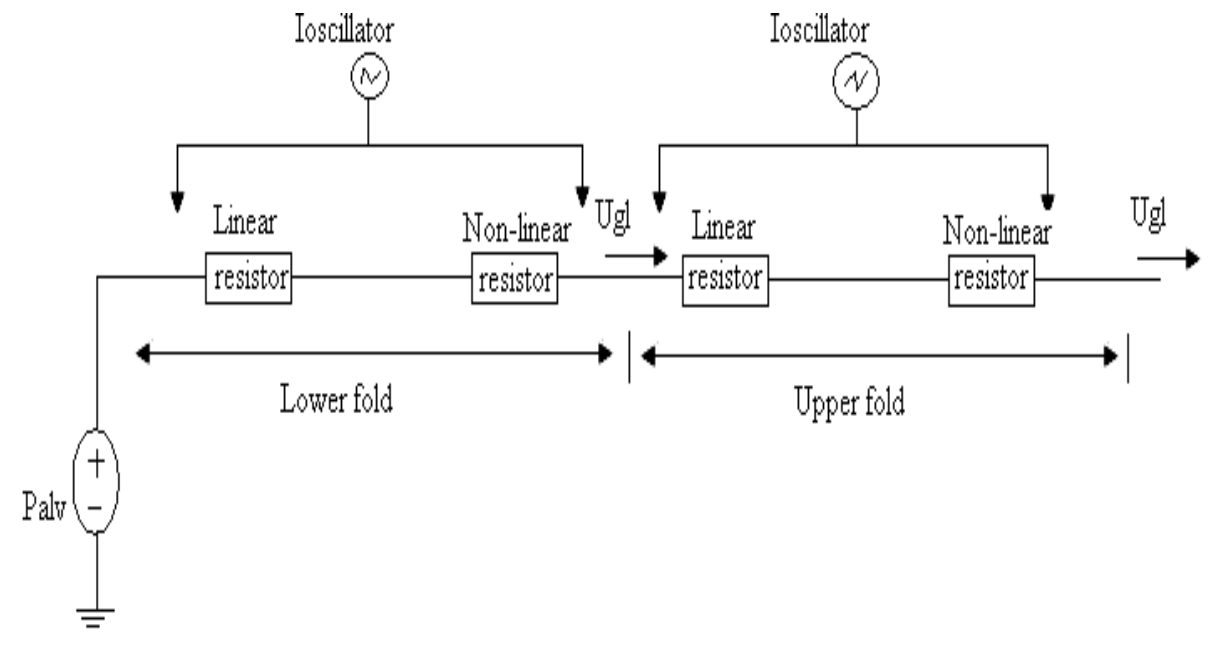

Fig.4 Complete glottal circuit

A normal voiced speech creates a pressure of about 7-10 cm $H_2O$ within lungs which in turn exerts force on vocal folds which make them to vibrate and in turn produces sound. To model this lung pressure, it is converted in dc voltage equivalent using equation (1), which acts as input voltage $P_{alv}$ for the glottal circuit. For normal voice, a nominal pressure of about 10 cm $H_2O$ is generated and correspondingly the dc equivalent voltage $P_{alv}$ is approximately 3 volt. The simulated output of glottal circuit is in form of periodic current pulses shown in Fig. 5, along with its derivative. The frequency range between successive glottal pulses is about 125Hz which satisfies the fundamental frequency generated during voiced speech in normal mode by male speakers.

According to the source-filter theory of speech production, lip radiation is represented by the derivative of the produced acoustic signal which means voice source is actually derivative of glottal pulse. Thus, the intensity of the produced acoustic wave depends rather on the derivative of the glottal flow signal than the amplitude of the flow itself. In other words, the derivative is the effective excitation of the vocal tract [11]. The principal acoustic excitation of vocal tract occurs at the discontinuity of derivative pulses. For analysis purpose, a single pulse of glottal circuit along with its derivative is taken, shown in Fig. 6. More the negative value of derivative, higher is the excitation. Generally, the inertia of air in the glottis and supraglottal airways prevent the occurrences of the abrupt discontinuities that occur at time of vocal fold closure. With the assumption of complete glottal closure, there will be discontinuity in first derivative at the endpoints of the open phase of the glottal wave. For voiced speech, the glottal flow derivative consists of two phases, that is, open phase and closed phase. During the open phase, vocal folds are progressively displaced from their initial state due to increasing subglottal pressure. When the elastic displacement limit of folds is reached, they suddenly return to the position of closing phase. There is slight time-gap between vocal folds separation, primarily due to the inertia of the vocal tract air below and above the glottis. Likewise, as the vocal folds come together during each oscillation, the inertia of the air supports and maintains a high flow until the closing of the glottis finally forces the flow to zero (assuming a total glottal closure).

During the most closed portion of the glottal cycle, where the flow is minimum; the waveform is relatively flat. For a normal voiced glottal cycle, there must be significantly long period in which the glottis is either closed, or sufficiently closed so that the glottal impedance is high enough to satisfy this condition. From simulation results, it can be observed that the glottal waveform admittance before and after the glottal pulse is zero. It is often desirable to monitor the degree of abduction or adduction of the vocal folds during voiced speech, both for steady voicing and during abductory or adductory movements.

## 4. Conclusion

The Glottal Source is an important component of voice as it can be considered as the excitation signal to the voice apparatus. Its modelling increases the parametric flexibility of the system and permits to transform voice characteristics of the speech. Using MOS model of glottis, drastic reduction of power consumption can be achieved which could be useful in portable speech processing systems of moderate complexity, like in cell phones, digital assistants, and laptops.

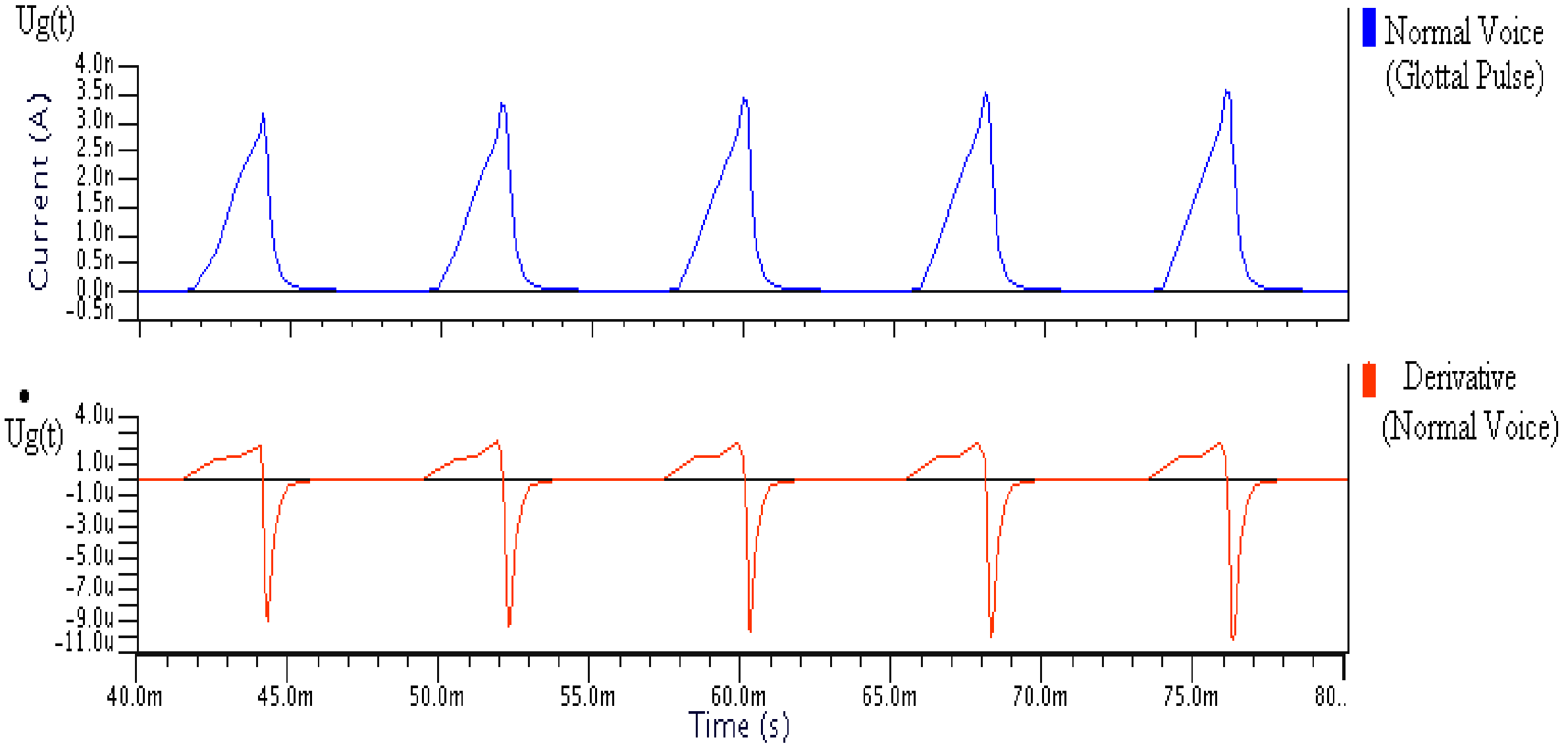

Fig. 5 Glottal pulse and its derivative for normal voice

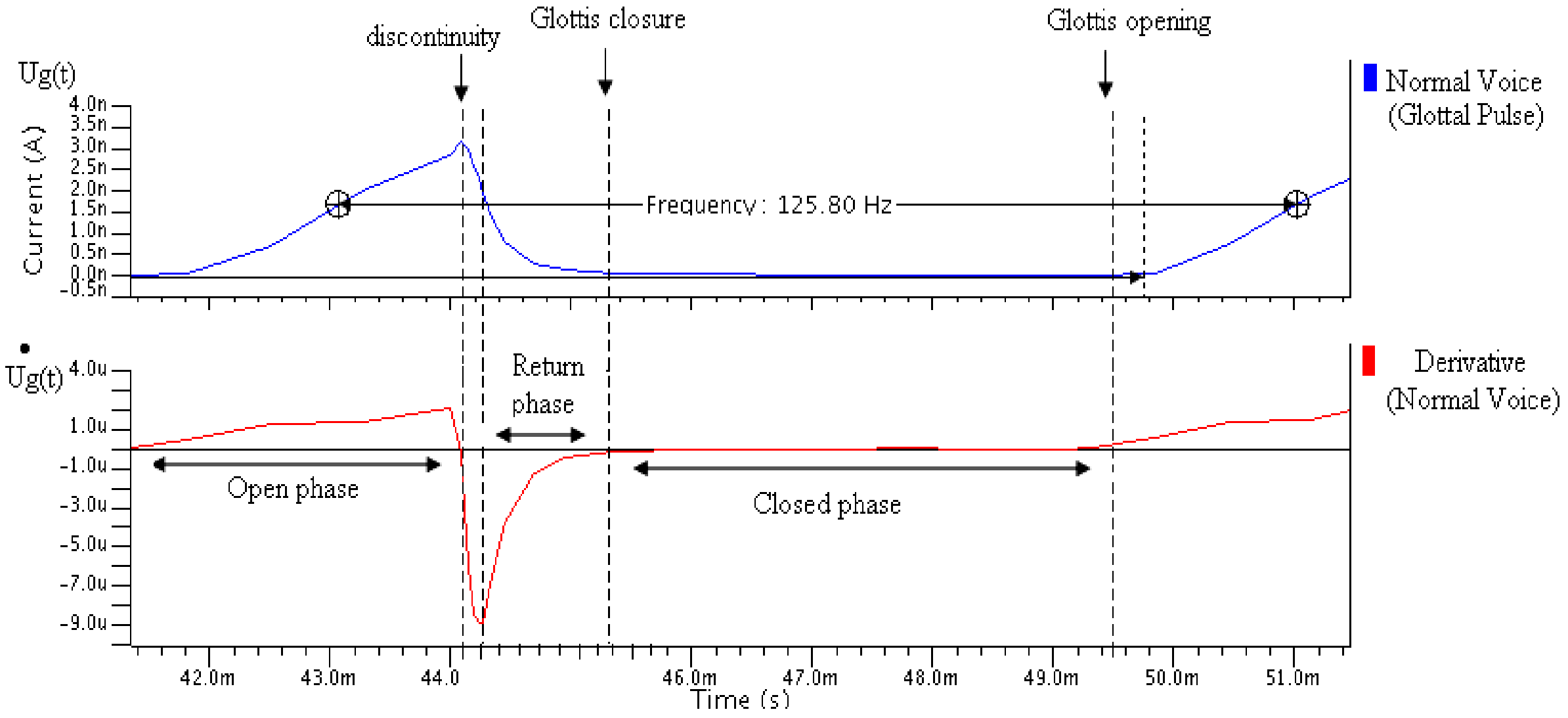

Fig.6 One single pulse of glottal wave and its derivative for normal voice

## Acknowledgments

The authors would like to thank to Mr. Wee, Mr. Turicchia and Mr. Sarpeshkar for their meaningful discussion on vocal tract model. The authors also extend their thanks to generous support of SMDP VLSI Design Lab of ECE department of NIT Kurukshetra and the financial assistance given by Special Manpower Development Programme (SMDP) project sponsored by ministry of communication and information technology, government of India.

**Nikhil Raj** received his M.Tech degree in Electronics and Communication Engineering with specialization in VLSI Design from National Institute of Technology Kurukshetra, Haryana, India in 2009. Currently, he is a lecturer with the Department of Electronics and Communication Engineering, NIT Kurukshetra, India. After doing many research works in VLSI area, he is currently doing research in bio-inspired analog circuits.

**R. K. Sharma** received his M.Tech in Electronics and Communication Engineering and Ph.D. degree in VLSI Design from National Institute of Technology Kurukshetra, Haryana, India in 1993 and 2007, respectively. Currently, he is a Associate Professor with the Department of Electronics and Communication Engineering, NIT Kurukshetra, India. His main research interests are in the field of low power VLSI design, electronic measurements, microprocessor and FPGA based measurement systems.